\documentclass{jdfsl}

\usepackage{graphicx}

\begin{document}
	
	\title{ 
		Essential Characteristics of Approximate matching algorithms \newline
		{\large A Survey of Practitioners Opinions and requirement regarding Approximate Matching}
	}

	\author{Monika Singh}
	
	\institute{Indraprastha Institute of Information Technology, Delhi (IIIT-D), Delhi, India
	}

	\abstract{
		
		Digital forensic investigation has become more challenging due to the rapid growth in the volume of encountered data. It is difficult for an investigator to examine the entire volume of encountered data manually.
		Approximate Matching algorithms are being used to serve the purpose by automatically filtering correlated and relevant data that an investigator needs to examine manually. Presently there are several prominent approximate matching tools and technique those are being used to assist critical investigation process.
		However, to measure the guarantees of a tool, it is important to understand the exact requirement of an investigator regarding these algorithms.  \\
		This paper presents the findings of a closed survey conducted among a highly experienced group of federal state and local law enforcement practitioners and researchers, aimed to understand the practitioner and researcher's opinion regarding approximate matching algorithms. The study provides the baseline attributes of approximate matching tools that a scheme should possess to meet the real requirement of an investigator.

	}
	
		\keywords{Digital Forensics, Approximate Matching Algorithm, Survey, Tools, Similarity Hashing}

	\maketitle

	\section{Introduction}\label{sec:introduction}
	
	In today's digital era, we are surrounded by enormous amount of digital data around us,  which could be computer hard-disc, external hard-drive, pen-drives, mobile storage, flash drives, tablets, etc. Hence on a crime scene, an investigator is confronted with several terabytes of digital data, which is an enormous volume of data for an investigator to examine manually. The major requirement of today's forensic investigation process is to have the capability to filter out the relevant data from the total volume obtained data on a crime scene that can be examined manually in a reasonable period of time. 
	
	Filtering is typically done by matching the case files with NIST reference data set. The Filtering process can be performed in following two ways 1) Blacklisting 2) Whitelisting. 	\textbf{Blacklisting} is the process of filtering data by matching them with the set of Known-to-be-bad files (as determined by the investigator). The resultant files after this process are the ones which an investigator needs to examine closely. \textbf{Whitelisting} is the process of filtering by matching the files with a set of already Known-to-be-good files. The files passing this process need not be examined by the investigator.

	Nowadays, Approximate Matching algorithms are being used to perform the filtering. Traditional cryptographic hash functions cannot be used to serve the purpose as even a single bit change in the file content is expected to modify the entire digest randomly, which is useful to find exact duplicates. However, here the requirement is to find similarity. 
	
	`Approximate Matching' is a generic technique for finding similarity among given files, typically by assigning a `similarity score.' An approximate matching technique can be characterized into one of the following categories: Bytewise Matching, Syntactic Matching, and Semantic Matching  
	Bytewise Matching measures the similarity at the byte level of the digital object without considering the internal structure of the data object. These techniques are known as fuzzy hashing or similarity hashing. Syntactic Matching defines similarity based on the internal structure of the data object. Semantic Matching measures similarity based on the contextual attributes of the digital objects. It is also known as Perceptual Hashing or Robust Hashing.

	Most prominent and commonly used approximate matching schemes includes ssdeep, sdhash, mrsh, etc. However, there are several cases where most of the existing scheme will fail to find similarity. For example similarity between a text document and PDF document with same content, colored and the grey scale version of the same image, different format (jpg, bmp, etc.) representation of the same image.
	The purpose of these algorithms is to assist the critical forensic investigation process by filtering out the relevant documents. Hence it is important to understand the real required properties that an algorithm should possess to filter out the actual relevant documents. 
	
	Hence we conducted a survey in a closed group of highly experienced and trained digital investigators, practitioners, and researchers to understand their requirements/needs regarding these algorithms.

	\section{Current Trends of Approximate Matching scheme}
	
	Approximate matching algorithms have been used since the year 2002. The first scheme was proposed by Nicolas Harbour \protect\cite{harbour2002dcfldd} called dcfldd, which is a block-based hashing scheme. Later an improvement of dcfldd was proposed by Kornblum~\protect\cite{kornblum2006} known as  Context Triggered Piecewise Hashing (CTPH). The CTPH scheme is based on an email detection algorithm called spamsum, proposed by Andrew et al.~\protect\cite{tridgell2002spamsum}.  Instead of hashing fixed size blocks, CTPH dived the data in variable size blocks and hashes each block using a non-cryptographic hash function called FNV hash. The CTPH tool is known as \textbf{ssdeep}. Breitinger et al. \protect\cite{breitingersecurity} presented the thorough analysis of ssdeep and showed that ssdeep does not withstand an active adversary for blacklisting and whitelisting.
	
	Roussev et al.\protect\cite{roussev} proposed a new scheme called sdhash in the year 2010. The basic idea of the sdhash scheme is to generate the final hash using only statistically improbable features of the document. Detailed security and implementation analysis of sdhash is presented in ~\protect\cite{breitingersecurity} by Breitinger et al. This work uncovered several implementation and security issues and showed that it is possible to beat the similarity score by tampering a given file without changing the perceptual behavior of this file (e.g., image files look almost same despite the tampering). The claims of \protect\cite{breitingersecurity} is again verified by chang et al. in \protect\cite{changcollision}. The paper also shows an attack method which can mislead the investigator with many forged similar files.
	Furthermore, Roussev\protect\cite{roussev2011evaluation} has shown that sdhash outperforms ssdeep in terms of both accuracy and scalability.
	
	Another scheme known as bbHash~\protect\cite{breitinger2012fuzzy} was proposed by Breitinger et al. in the year 2012. However because of the high runtime bbHash is not practically usable. Breitinger et al. proposed another scheme mvHash-B similarity preserving hashing ~\protect\cite{mvhash} in year
	2013 (Breitinger et al., 2013). The scheme works in three phases; first, compresses the input data using majority voting, performs run-length encoding and then finally stores the fingerprint into Bloom filters. `B' in mvHash-B denotes the bloom filter representation of the similarity digest. In terms of performance, mvHash-B is one of the most efficient schemes among all existing schemes with lowest run-time complexity and small digest size. A thorough analysis of mvHash-B is presented by Chang et al. ~\protect\cite{chang2016security}. The paper uncovers the weakness of mvHash-B scheme and shows that mvHash-B does not withstand an active adversary against the blacklist and also proposes an improvement to mvHash-B design to conquer the weakness. 
	
	Apart from the above-mentioned schemes, several forensic tools are using different filtering techniques. FTK performs Cluster Analysis to find related documents, near duplicates. Encase uses Entropy Near Match Analyser to discover similar files. X-ways implemented a new technology called FuzZyDoc to identify known documents.


	
	\section{Survey Methodology}
	\subsection{Purpose Of the Survey}
	
	The baseline definition and terminology of the approximate matching algorithm is already defined by \protect\cite{breitinger2014approximate}. 
	The NIST Special Publication 800-168 \protect\cite{breitinger2014approximate} defines the properties at more general and broader level. However, similarity definition, as well as the requirements, vary for different data object type. For example, files with similar perceived text content may have entirely different structure and would be entirely different if an inappropriate algorithm is applied; the actual similarities would remain unnoticed. A color and the grayscale version of the same image would be completely different when using most of the existing schemes.  Hence there is a strong need to define the properties based on the practical requirements. 
	
	The aim of the survey was to establish the key characteristics of the approximate matching algorithm based on the requirement of digital forensics practitioners and researchers with the understanding of different perspectives towards these algorithms. Using the results of the survey, in future, our aim is to build the real data-set with known similarity and evaluate the existing schemes based on the key characteristics using the real data-set.

	\subsection{Survey Design}
	The survey contains a brief introduction to the topic of the study and the purpose of the survey, followed by 10 questions consisting :
	
	\begin{itemize}
		\item 4 optional questions
		\item 1 Multiple choice question
		\item 4 likert scale questions
		\item 1 Ranking question
	\end{itemize}
	
	Each of the questions contained one comment section to share any other information regarding the question, which allowed the participants to express their view in more pragmatic and reasonable manner. 
	
	The format of the survey was kept simple and short while capturing all the information needed, In order to maintain the balance between the amount of time and effort required to complete the survey. The survey was conducted online among a closed group of 80 digital forensic practitioners; those are part of NIST steering committee and familiar with the use of NSRL for filtering. The audience of the survey was kept limited in-order to assure the reliability and legitimacy of the results. The survey was conducted in the entirely anonymous manner, no personally identifiable information about the participants was stored. It was available online for two weeks. Among the 80 invitation, we received total 19 completed responses, which is an acceptable amount of data to analyze and conclude the results of the survey for such highly targeted surveys.

	\subsection{Survey Results}
	The results of survey is presented in this section based on the all the received responses. However, where appropriate, further analysis based upon the perspectives of researchers and practitioners is presented.

	\subsubsection{Demographic Information}
	
\begin{figure}[ht!]
		\includegraphics[scale=0.64]{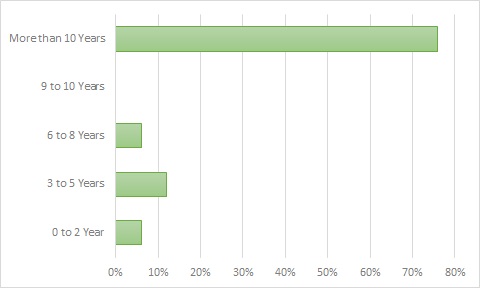}
		\caption{Demographic Data}
		\label{figrle}
	\end{figure}

	Demographic Information of the survey is represented in figure~\ref{figrle}. More than 75\% of the participants had more than 10 years experience as shown in figure~\ref{figrle}. All of the participants were from the United States of America and working as a federal, state or local law enforcement practitioner or researchers.

	\subsubsection{Awareness About Approximate Matching Algorithms}
	
	Results shown in around 65\% of the total participants were aware and are using approximate matching algorithms during the investigation process. Among them, 45\% were aware that their tools perform approximate filtering but were unaware of the technique/scheme used by the tool. Whereas remaining were well aware of the techniques used by the tools or the specific technique used by them. Following is the list of schemes mentioned by the participants those are being used by the practitioners these days: Cluster analysis of FTK, FuzzyDocs of X-ways, md5Deep, CodeSuite, PhotoDNA, sdhash, ssdeep and some of the practitioners are using their self-constructed tf-idf based schemes.
	This shows that most of the practitioners find approximate matching algorithms useful and are willing to use these algorithms to filter out relevant data. However, since there is no consensus for one standard scheme hence all of them are using techniques that are either by default being used by the tools or they find it more useful for a particular case based on their personal experience.

	\subsubsection{Primary Applications of Approximate Matching Algorithms} 
\begin{figure}[ht!]
		\includegraphics[scale=0.64]{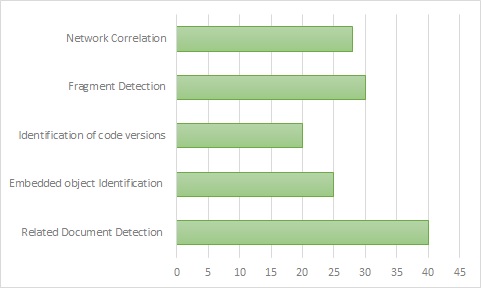}	
		\caption{Applications of Approximate Matching Algorithms}
		\label{fig:uses}
	\end{figure}

	Participants were asked to scale the uses of the approximate matching algorithm based on their professional experiences. List of applications was derived by the examining the existing literature shown in figure~\ref{fig:uses}. Responders were also allowed to add more uses to the list.
	
	\begin{itemize}
		\item Based on all of the received responses approximate matching algorithms are being frequently used to identify the Related Documents.
		\item  Fragment Identification was scaled second in the list. Fragment Identification is the identification of a document based on a piece of data.
		\item Following  are the other important uses in the respective order of their raking
		\begin{itemize}
			\item Correlation of network (Data packet reconstruction from the fragmented files over the network)
			\item Embedded Object Identification ( e.g., a jpeg within a word document)
			\item Identification of the code version (Identification of patched or upgraded version of software).

		\end{itemize}

		However approximate matching algorithms don't work well for network correlation if the network is encrypted.

	\end{itemize}

	
	On further examination about the type of the data object that they need to filter out in most frequent cases, the participants were asked to rate among text, image, executable file, and multimedia. All of the received responses shows that text and images are the most commonly appeared data types that need to be filtered. Executable and multimedia files also required to be filtered in some of the cases.
	
	Hence we can state that one of the important characteristics of a scheme is its ability of related document correlation and fragment identification for text and image data type.

	\subsubsection{Key Measure to Identify the Ground Truth}
	
\begin{figure}[ht!]
		\includegraphics[scale=0.60]{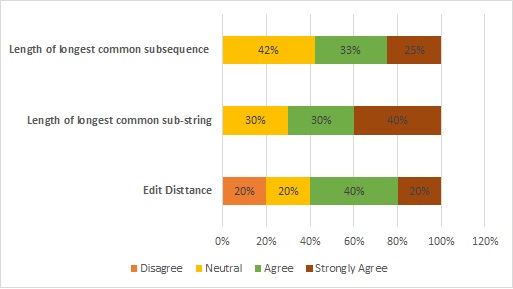}	
		\caption{Key Measure to Identify the Ground Truth}
		\label{fig:Text}
		
	\end{figure}

	Participants were asked choose among the existing measure that represents the similarity of two textual documents most closely in their opinion. Proposed measures were `Edit distance,' `Length of longest common sub-string' and `Length of longest common subsequence.' Where Edit Distance is a way of calculating dissimilarity between two text sequence by counting the minimum number operations required to transform one sequence into the other. Length of longest common substring denotes the length of the common contiguous substring of maximal length. Length of the longest common subsequence of two text sequence signifies the length of the common substring of maximal length where the substring might not appear in contiguous fashion but preserves the ordering of characters. All of these measures were taken by examining the current literature. Participants were also allowed to add new measures.  The purpose of this question is to find out the key measure to find real similarity in two text documents. Figure~\ref{fig:Text} shows that based on the experience of the responder `Length of longest Common Substring' defines similarity between two documents the most. Hence this measure should be used to build ground truth for Text data set with known similarity.

	However by further analysis of the responses, we realized this should be analyzed further to make a persuasive conclusion.

	\subsubsection{Image Similarity}
	
\begin{figure}[ht!]
		\includegraphics[scale=0.53]{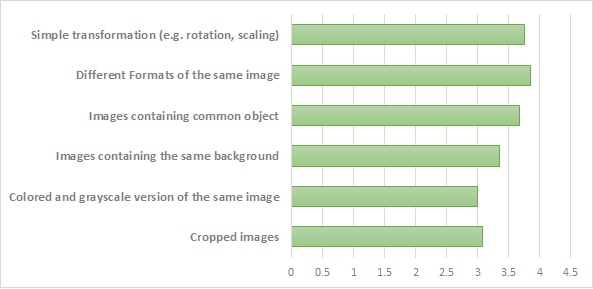}	
		\caption{Image Similarity}
		\label{fig:imagee}
	\end{figure}

	Participants were asked to rank the similar images that they need to filter out during the real case investigations. Figure~\ref{fig:imagee} shows that all of the listed image similarities are almost equally important from the practitioner's point of view. However, the ability of a scheme to find out the similarity between different formate is ranked highest. Hence it would be useful if a technique can filter out all of the listed image similarities.   
	
	\subsubsection{Executable Program File Similarity}
\begin{figure}[ht!]
		\includegraphics[scale=0.60]{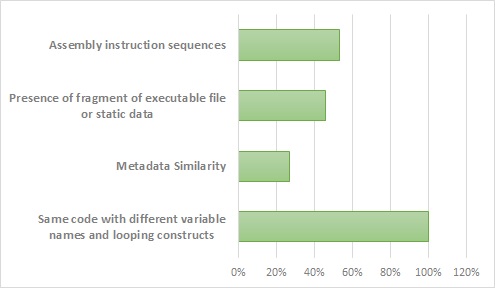}	
		\caption{Executable Program File Similarity}
		\label{fig:executable}
	\end{figure}

	The purpose of this question to understand similarity definition for a program file based on the requirement of an investigator. Figure~\ref{fig:executable} shows the results. All of the listed similarity characteristics are chosen by examining the current literature and participants were allowed to add missing properties. From the received responses we can say one of the essential abilities of a tool for program file similarity is to find the same program with different variable name and looping constructs. According to practitioner most of the existing tools produce high false positive results because of the flagging reusable components such as dlls, icons and other resources obtained from software libraries.
	
	\subsubsection{File System Information} Participants were asked that if the volume data volume obtained from the crime scene containing file structure information such as MFT(Master File Table)  for windows files, inode, etc.  has increased. From the result, we can say more than 70\% of the cases the file structure information is available. Hence this information can be used to improve the filtering abilities of the tools. We also found that amount of application specific data has also increased such as SQLite data on mobile devices. Hence major challenge for approximate matching algorithms is to have the ability to find similarity across the device specific data.
	
	\begin{figure}
		\includegraphics[scale=0.617]{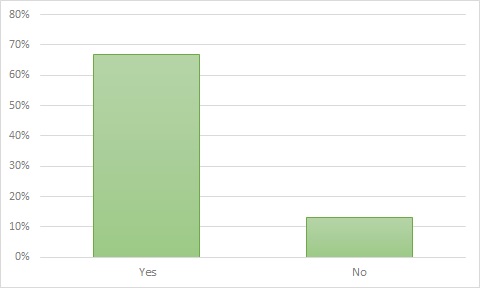}	
		\label{fig:image}
		\caption{File System Information}
	\end{figure}

	\section{Conclusions and Future Work}
	
		Automated filtering has become one of the essential requirements of today's digital investigation process. Approximate matching algorithms are being used to perform the filtering. Approximate matching is a relatively new area of digital forensics, which is still evolving and needs to be defined more formally. The aim of the survey was to gain insights of the practitioners and researchers opinion regarding this new class of algorithms. From the results of the study, we found that practitioners and researchers acknowledge the utility of approximate matching algorithms and are willing to use it to speed up the investigation process. Since there is no standard way to measure the guarantees of the algorithms. It is difficult to find the appropriate tools.
	
	Hence this paper presents the practitioners and researcher's requirement and perspective towards approximate matching algorithms. Future work must focus on formalizing the properties of approximate matching algorithms and providing a well-defined evaluation framework to evaluate existing and future approximate matching schemes.

	
	\bibliographystyle{apacite}
	\begin{flushleft}
		\bibliography{literature}
	\end{flushleft}
	
\end{document}